\documentclass[aps,groupedaddress,showpacs]{revtex4}
\usepackage{graphicx}
\newcommand{\figwidth}{10cm}
\begin{document}
\newcommand{\cm}{cm$^{-1}$}
\title{Orbital Dimerization and Dynamic Jahn-Teller Effect in
NaTiSi$_2$O$_6$}
\author{M. J. Konstantinovi\'c
$^{a,*}$, J. van den Brink, $^{b}$ Z. V. Popovi\'c $^{a,c,**}$, V.
V. Moshchalkov $^{a}$, M. Isobe $^{d}$, and Y. Ueda $^{d}$}

\affiliation{ $^a$ Laboratorium voor Vaste-Stoffysica en
Magnetisme, Katholieke Universiteit Leuven, Celestijnenlaan 200D,
B-3001 Leuven, Belgium }
\affiliation{$^b$ Institut-Lorentz for
Theoretical Physics, Universiteit Leiden, P.O. Box 9506, 2300 RA
Leiden and
Faculty of Applied Physics, University of Twente, P.O. Box 217, 7500 AE
Enschede, The Netherlands
}
\affiliation{$^c$ Institute of
Physics-Belgrade, P. O. Box 68, 11080 Belgrade/Zemun,
Yugoslavia}
\affiliation{$^d$ Institute for Solid State Physics,
The University of Tokyo, 5-1-5 Kashiwanoha, Kashiwa, Chiba
277-8581, Japan}

\begin{abstract}
We study with Raman scattering technique two types of phase
transitions in the pyroxene compounds NaMSi$_2$O$_6$ (with M=Ti,
V, and Cr). In the quasi one-dimensional S=1/2 system
NaTiSi$_2$O$_6$ we observe anomalous high-temperature phonon
broadening and large changes of the phonon energies and
line-widths across the phase transition at 210 K. The phonon
anomalies originate from an orbital order-disorder phase
transition and these results --combined with theoretical
considerations-- indicate that the high temperature dynamical
Jahn-Teller phase of NaTiSi$_2$O$_6$ exhibits a spontaneous
breaking of translational symmetry into a dimerized, Jahn-Teller
distorted, orbital ordered state under the formation of spin
valence bonds. In S=1 NaVSi$_2$O$_6$ orbital degrees of freedom
are strongly suppressed and the magnetic excitations are well
described within a Heisenberg model, indicating that at T$_N$=19K
this system orders antiferromagnetically.
\end{abstract}

 \maketitle

{\it Introduction.}
Electrons in strongly correlated transition-metal compounds can be
regarded as having separate spin, charge and orbital degrees of freedom.
It is the interplay between these, combined with their coupling to the
lattice, that gives rise to a wealth of possible spin, charge and
orbital orderings, as observed for instance in many Colossal
Magneto-Resistance manganites~\cite{a1}, cuprates~\cite{Kugel},
titanates (e.g. LaTiO$_3$~\cite{Keimer}) and vanadates
(e.g. V$_2$O$_3$~\cite{Castellani} and LiVO$_2$~\cite{Pen}).

Systems with orbital degeneracy are particularly interesting because
orbitals couple to the lattice via the cooperative Jahn-Teller
(JT) effect on one hand, and via superexchange interactions to the
electronic spin on the other hand~\cite{Kugel}. Therefore at an orbital
ordering phase transition the magnetic susceptibility and phonon
properties will be affected at the same time. Experimentally, however,
such an interrelation is seldom found: in general the dominant JT
orbital-lattice coupling obscures the more subtle effects due to
the superexchange.

Pyroxenes are large family of compounds (AMB$_2$O$_6$;
A=alkali-metal, M=transition-metal, and B=Si, Ge) which structure
consists of isolated quasi one-dimensional chains of edge-sharing
MO$_6$ octahedra, linked together by the corner-sharing BO$_4$
tetrahedra, Fig. 1. Particularly, the compounds with {\it active}
$t_{2g}$ orbitals (M=Ti,V, and Cr) are expected to have electronic
interactions governed by both orbital degeneracy and anisotropy.
Interestingly, the sodium-silicon system with titanium,
NaTiSi$_2$O$_6$, is rather different from other pyroxenes as it
lacks low-temperature AF order, and shows signs of the opening of
a spin gap instead \cite{a2,a3,a30,a31}. Ti$^{3+}$ corresponds to
S=1/2, and since all Ti ions are in equivalent crystallographic
positions, this material is a prime candidate to show a
spin-Peierls (SP) phase transition. Indeed, its magnetic
susceptibility~\cite{a2} sharply decreases below 210 K, indicating
a transition to a spin-singlet state, which could in principle be
of the SP type. But already the high-temperature magnetic
susceptibility data does not support such a SP scenario, since the
phase transition occurs at a temperature that is higher then the
maximum point of the Bonner-Fisher curve, which implies that the
transition cannot solely be driven by magnetic
fluctuations~\cite{a2}. Based on structural analysis, Isobe et.al.
anticipated that orbital dimerization might be responsible for
such behavior \cite{a2}.

In this Letter, we report a Raman scattering study on a family of
pyroxene compounds from which we conclude that orbital degrees of
freedom play decisive role in physical properties of
NaTiSi$_2$O$_6$, and that NaTiSi$_2$O$_6$ undergoes an orbital
order phase transition at T$_{\rm OO}$=210 K. At high temperatures
the fluctuations of the orbital degrees of freedom lead to a
dynamic Jahn-Teller phase with anomalous phonon broadening and
remnant antiferromagnetic (AF) spin fluctuations. We find that the
dramatic drop of the magnetic susceptibility below T$_{\rm OO}$ is
accompanied with a structural change, just as one would expect for
a canonical orbital ordering transition. These results, combined
with the microscopic orbital-spin model that we derive, suggest
that the quasi one-dimensional dynamical Jahn-Teller phase of
NaTiSi$_2$O$_6$ exhibits a spontaneous breaking of the
translational symmetry into an dimerized orbital ordered state
with a spin gap due to the formation of spin valence bonds. The
paper is organized as follows: first we present the Raman spectra
and underline important observations, then we discuss the
electronic structure of pyroxenes and derive a microscopic model.
Finally, we demonstrate that all observed effects (including
susceptibility data) can be accounted by our model.

\begin{figure}
\includegraphics[width=\figwidth]{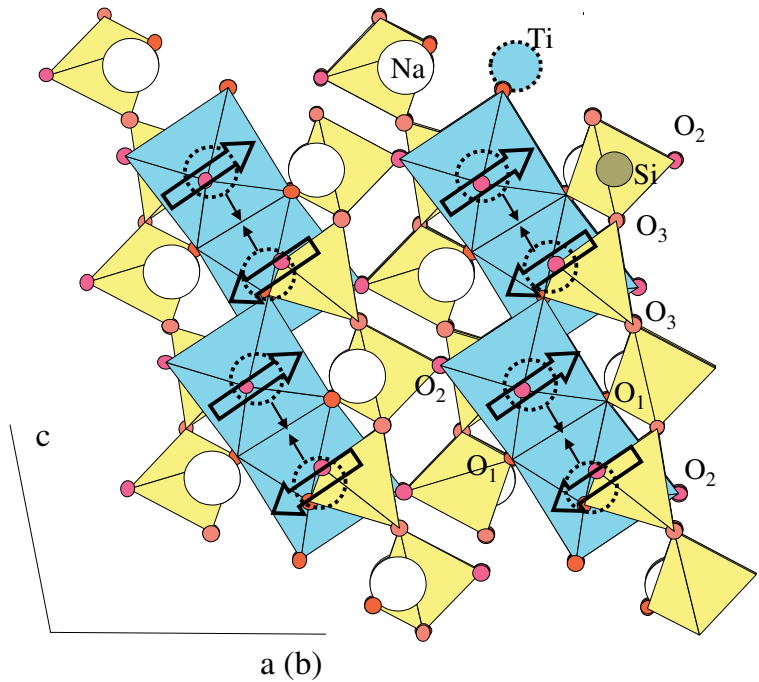}
\vspace{-1em}
\caption {a) Crystal structure of NaTiSi$_2$O$_6$. Block arrows
denote magnetic moments, and small arrows represent possible
distortions (dimerization) of the TiO$_6$ octahedra in the low
temperature phase.  \label{fig1}}
\end{figure}

{\it Experiment.}High-quality powder samples of NaMSi$_2$O$_6$
(M=Ti, V, and Cr) were prepared by a solid-state reaction of
mixtures with an appropriate molar ratio of Na$_2$MSi$_4$O$_{11}$,
M and MO$_2$. Details of the sample preparation are published
elsewhere \cite{a2}. Raman spectra are measured in the
backscattering configuration, using 514.5 nm line of an Ar-ion
laser, micro-Raman system with a DILOR triple monochromator, and a
liquid nitrogen cooled charge-coupled device detector. For the
low-temperature measurements we used the Oxford continuous-flow
cryostat with a 0.5 mm thick window. The focusing of the laser
beam is realized with a long distance (10 mm focal length)
microscope objective (magnification 50$\times$).

The pyroxenes crystallize in a monoclinic unit cell with the space
group $C2/c$ \cite{a4}. The unit cell consists of four formula
units (Z=4) with 40 atoms in all. The site symmetry of Na, M, Si,
O$_1$, O$_2$ and O$_3$ atoms are (4e), (4e), (8f), (8f), (8f) and
(8f), respectively. Thus, the factor-group-analysis (FGA) yields:
(Na, M)($C_{2}$) $\Gamma$ = $A_g$ + $A_{u}$ + 2$B_{1g}$ +
2$B_{u}$; (Si, O$_1$, O$_2$, O$_3$)($C_1$) $\Gamma$ = 3$A_g$ +
3$A_u$ + 3$B_{g}$ + 3$B_{u}$. Summarizing these representations
and subtracting the acoustic modes ($A_{u}$ + 2$B_{u}$), we obtain
the following irreducible representations of NaMSi$_2$O$_6$
vibrational modes:
\begin{eqnarray*}
\Gamma_{NaMSi_2O_6}^{opt.} &=& 14A_g(xx,yy,zz,xz) + 16B_{g}(xy, yz) \nonumber \\
&+& 13A_{u}({\bf E}||{\bf y}) + 14B_{u}({\bf E}||{\bf x}, {\bf
E}||{\bf z})
\end{eqnarray*}
The unpolarized Raman spectra of NaMSi$_2$O$_6$ (M=Ti, V, Cr) are
shown in Fig. 2. At the room temperature we find around 30 phonon
modes as predicted by FGA. The Raman spectra of different
compounds are similar as expected for isostructural materials, and
the phonon modes can be crudely grouped into two energy regions.
The modes in the spectral range below 500 cm$^{-1}$ originate from
the bond bending vibrations, whereas the higher frequency modes
originate from the stretching vibrations. The highest energy modes
are mainly due to the non-bridging Si-O ion vibrations, because of
the shortest Si-O tetrahedral bonds. The modes of NaMSi2O6 at
1055/1032 cm$^{-1}$ (Cr), 1042/1025 cm$^{-1}$ (V), and 1042/1025
cm$^{-1}$ (Ti), represent Si-O$_2$ (see Fig. 2)
antisymmetric/symmetric bond stretching vibrations, respectively.
Their frequency difference scales as R$^{-3}$ in a full accordance
with difference between Si-O$_2$ bond lengths in these materials.
Similarly, the modes at 990/967 {\cm} (Cr), 972/954 {\cm} (V), and
965 {\cm}(Ti), we assign as antisymmetric/symmetric pairs of
Si-O$_1$ bond stretching modes. This simplified mode assignment
becomes inapplicable at lower frequencies due to more complicated
normal coordinates of corresponding vibrations. However, besides
similarities, we also observe two very important effects. First,
most of {\it the phonon line widths are dramatically increasing in
the components with smaller spin values} ($V^{3+}\rightarrow S=1,
Cr^{3+}\rightarrow S=3/2$). Second, the spectra of
NaTiSi$_2$O$_6$, due to the large phonon broadening, show
effectively less phonon modes then expected by FGA, and observed
in other pyroxenes. The latter effect may also be regarded as a
consequence of a high temperature {\it "higher symmetry lattice
state" of NaTiSi$_2$O$_6$}.

\begin{figure}
\includegraphics[width=\figwidth]{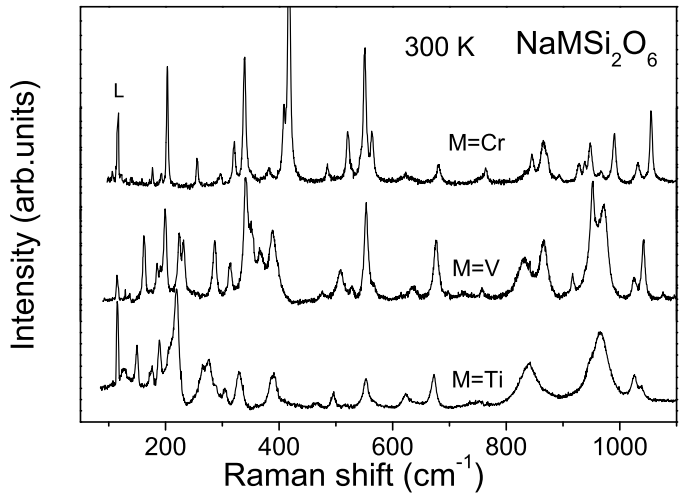}
\caption {Raman spectra of NaMSi$_2$O$_6$. \label{fig2}}
\end{figure}

By lowering the temperature we find a dramatic change in the
phonon Raman spectra of NaTiSi$_2$O$_6$ around 210 K, see Fig.
3a,b. The modes at 221 and 209 {\cm} exhibit the anti-crossing
behavior, see inset Fig.3a, which indicates that these two modes
belong to the same symmetry class below $T_c$. A similar effect is
observed for all other such pair of modes (one more pair is also
visible in Fig. 3a at about 180 and 190 {\cm}) in the spectra,
which suggests that in the low-T phase of NaTiSi$_2$O$_6$ all
Raman active modes are of the same (A$_g$) symmetry type.
Furthermore, our reflectance measurements (will be published
separately, Ref. \cite{a6}) show that the infrared active phonon
modes can be distinguished from the Raman modes, since the center
of inversion remains to be the symmetry element in the low-T
phase. Accordingly, we uniquely determine the space group of the
low-T phase of NaTiSi$_2$O$_6$ to be P$\overline{1}$. {\it This
space group requires the translation symmetry breaking of the
TiO$_6$ chain} (the tentative distortion pattern of the low-T
phase is shown in Fig. 1) which is in agreement with preliminary
X-ray diffraction and neutron scattering data \cite{a7}.

Moreover, we show in details the temperature dependence of the
structure around 970 {\cm}, Fig. 3b. The mode at about 946 {\cm}
softens by about 10 {\cm}, while mode at 966 {\cm} "splits", and
hardens by 25 {\cm}, see inset in Fig. 3b. The full width at half
maximum (FWHM) of the 946 {\cm} (circles) and 966 {\cm} (squares)
phonons (the FWHM is estimated from Lorenzian-fit) increase up to
the maximum value at about 210 K, and then decreases to the
saturation value which is much smaller then the T=300 K value, see
right inset of Fig. 3b. In fact, all Raman-active phonons exhibit
anomalies at the temperature which coincides with $T_c$ obtained
from susceptibility measurements \cite{a2} indicating that {\it
the magnetic ordering is accompanied with the structural phase
transition} just as one would expect for an canonical
orbital-ordering transition. Besides, behavior of FWHM implies
that {\it the bond fluctuations are considerably larger in the
high-T then in low-T phase}(proximity of the structural phase
transition induces the largest fluctuations, producing the maximum
FWHM at T$_{\rm OO}$). Due to the JT effect bond fluctuations
emphasize the strong orbital character of disorder above the phase
transition temperature\cite{a771,a772}.

\begin{figure}
\includegraphics[width=9cm]{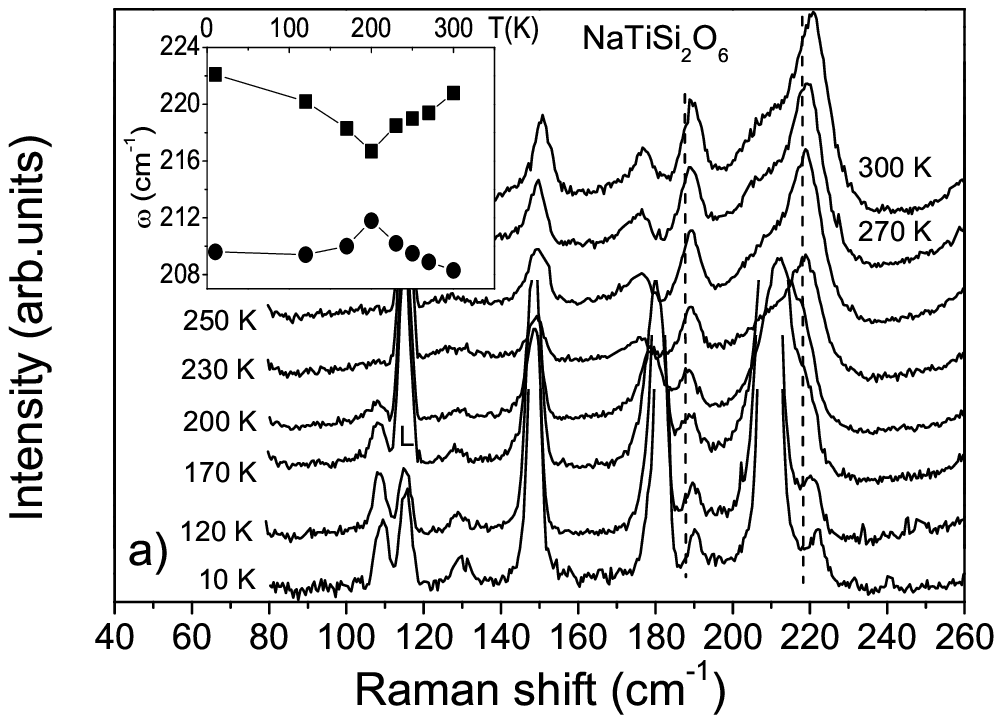}
\end{figure}
\begin{figure}
\includegraphics[width=9cm]{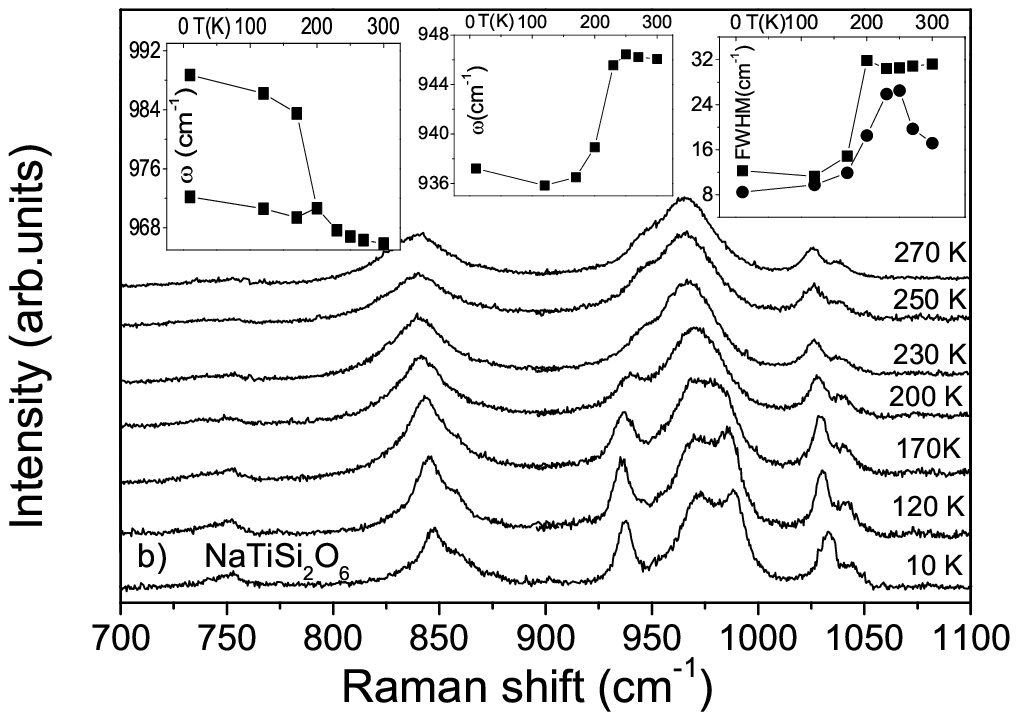}
\end{figure}
\begin{figure}
\includegraphics[width=9cm]{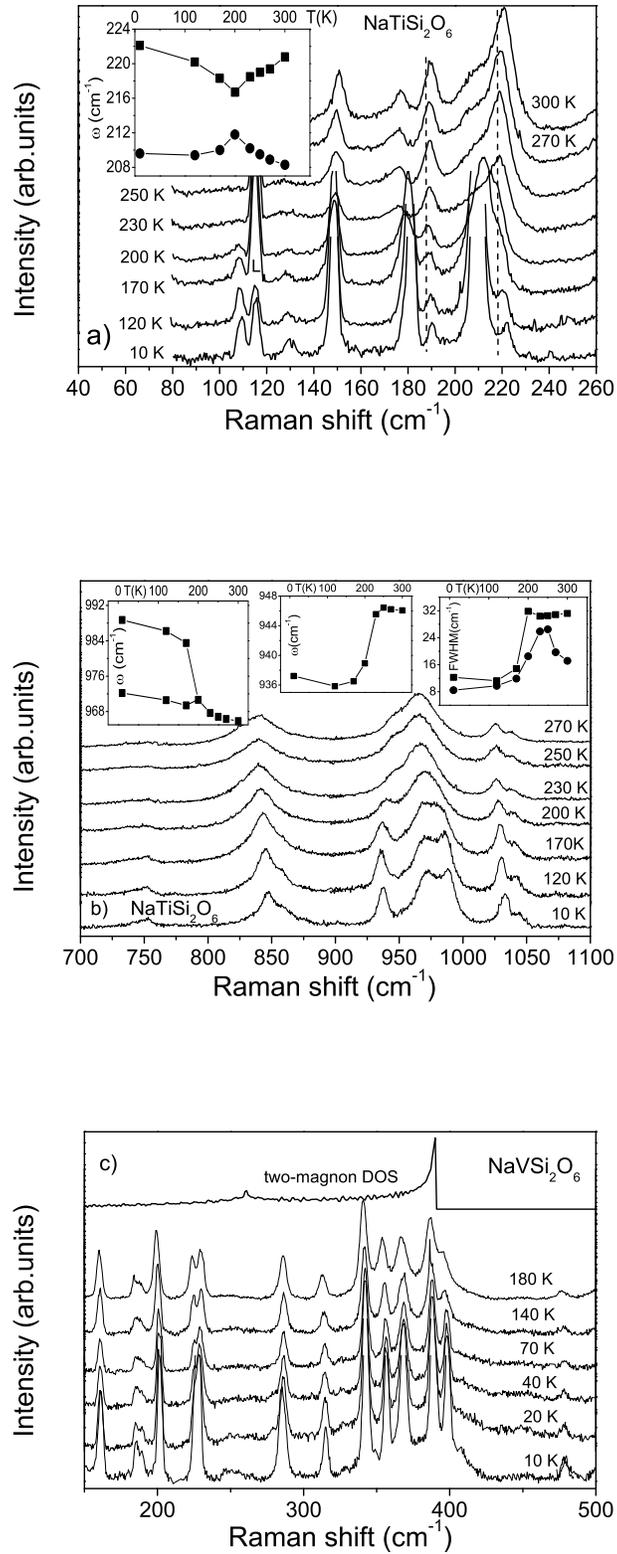}
\caption {Temperature dependence of the Raman spectra in a),b)
NaTiSi$_2$O$_6$, and c) NaVSi$_2$O$_6$. Top panel in c) shows
calculated two-magnon density of states. Insets: Temperature
dependencies of the frequency and line width of various phonons in
NaTiSi$_2$O$_6$. Vertical dashed line is an guide for the eye.
\label{fig3}}
\end{figure}

Contrary to S=1/2 compound, in the low temperature Raman spectra
of NaVSi$_2$O$_6$, see Fig. 3c, we did not observe the phonon
anomalies. Instead, at low temperatures, broad asymmetric
features, typical for the two-magnon excitations in Heisenberg AF
($H=\sum_{i,j}J_{i,j}{\bf S}_i \cdot {\bf S}_j, J_{i,j} > 0$),
appear around 260 and 400 {\cm}. According to the linear
Anderson's approximation, differential cross section for the
two-magnon (tm) Raman scattering is proportional to the two-magnon
density of states (DOS), $I_{tm}\sim \sum_{\bf k}\delta (\omega -
2\omega({\bf k}))$. The dispersion relation is obtained assuming
two-dimensional magnetic structure with different exchange
interactions along the chains (J$_{\parallel}$), and perpendicular
(J$_{\perp}$) to the chains: $\omega({\bf
k})=2S\sqrt{(J_{\parallel}+J_{\perp})^2-(J_{\parallel}cos(ka)+J_{\perp}cos(kb))^2}$.
The calculated two-magnon DOS is shown in top panel of Fig. 3c.
Two singularities at energies of about 260 and 390 {\cm} are
obtained for J$_{\parallel}=85$ {\cm} and J$_{\perp}=13$ {\cm}
(120 K and 18 K, respectively), in a very good agreement with
experiment. The maximum of the susceptibility curve in
NaVSi$_2$O$_6$ is at about 100 K \cite{a7} (giving $J_{\parallel}
\sim$ 80 K \cite{a301}) which is in good agreement with our
estimation of the J$_{\parallel}$. Thus, we conclude that {\it
NaVSi$_2$O$_6$ exhibits an AF ordering}. Similar AF phase has been
already observed in other S=1 chain compounds LiVGe$_2$O$_6$
\cite{a3}, and NaVGe$_2$O$_6$ {\cite{a31}.

{\it Pyroxenes: theoretical.} Next we discuss the electronic
structure. For a single octahedron the cubic crystal field splits
the Ti,V or Cr 3d-states into low lying $t_{2g}$ states, with 1,2
or 3 electrons, respectively, and empty states of $e_g$ symmetry
at higher energy. The low energy electronic properties are
governed by the three-fold degenerate $t_{2g}$ states (the
relevant states are $| xy \rangle$, $| yz \rangle $ and $| zx
\rangle$)~\cite{a701}.

The Coulomb interaction $U$ between electrons on the same
transition metal atoms is large, so the exchange interactions can
be determined by a second order perturbation expansion in the
electron hopping parameters. The problem is further reduced by
considering the symmetry allowed hopping paths in the chain
geometry, schematically represented in Fig. 4. If we consider
orbitals on two sites in the same $xy$ plane, then only the
hopping between $| xy \rangle$ orbitals is relevant (see Fig. 4a).
For sites in the $yz$ plane, the $|yz \rangle$ orbitals are
relevant, Fig. 4b. In the present geometry (note that x, y, and z
in Fig. 4 do not correspond to a, b, and c axis in Fig. 1) there
are no transition metal atoms in the same chain that are also in
the same $xz$ plane.  The $| xz \rangle$ orbitals are therefore
non-bonding and can be considered inert on this level of
approximation (a tightbinding parametrization~\cite{a5a} shows
that other overlap integrals either vanish by symmetry or are more
than factor five smaller), see Fig. 4c. For the S=1/2 Titanium
system we then obtain the Hamiltonian~\cite{a5b},
\begin{equation}
H^{\rm Ti} =\mid {\rm J^{Ti}} \mid \sum_{i,j}{\bf S}_i \cdot {\bf
S}_j [ \frac{1}{4}+T_i^z T_j^z + \frac{(-1)^i}{2}(T_i^z+T_j^z) ],
\label{eq:ham}
\end{equation}
where we use the orbital operators $T$ ($T^z_i=1/2$ corresponds to
an occupied $| xy \rangle$ orbital and $T^z_i=-1/2$ to an occupied
$| yz \rangle$ orbital on site $i$), and $i,j$ are neighboring
sites. For the S=1 Vanadium and S=3/2 Chromium system the orbital
degree of freedom vanishes ($T=0$) as the both of the {\it active}
orbitals are occupied, so for the spin system we obtain a simple
Heisenberg Hamiltonian $H^{\rm V/Cr}= \mid {\rm J^{V/Cr}} \mid
\sum_{i,j}{\bf S}_i \cdot {\bf S}_j$. This immediately explains
the uniqueness of the Ti (S=1/2) system, and the
antiferromagnetism in NaVSi$_2$O$_6$ (in fact in all S $\neq 1/2$
pyroxenes).

\begin{figure}
\includegraphics[width=\figwidth]{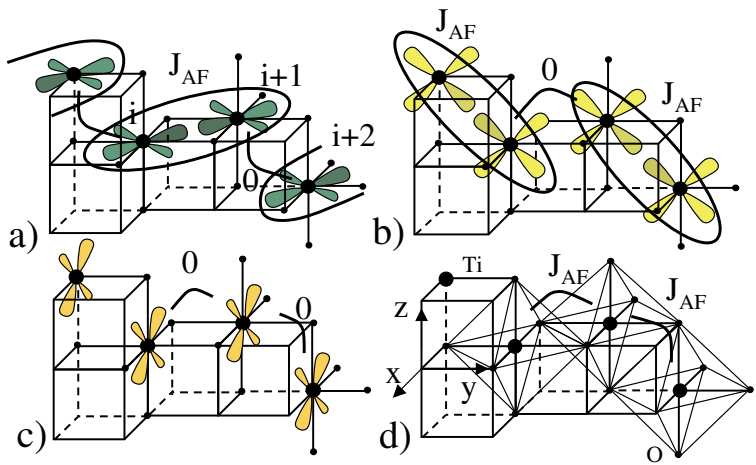}
\vspace{-1em}
\caption { Schematic presentation of the orbital ordering in
NaTiSi$_2$O$_6$. Orbital overlap in a) xy, b) yz, and c) xz
planes. d) Heisenberg AF phase. \label{fig4}}
\end{figure}

{\it Interpretation of experiments.} The ground state of
Hamiltonian~(\ref{eq:ham}), is clearly a {\it ferro-orbital}
state, with spin-singlets on each bond, where the energy per dimer
is $-3 {\rm J^{Ti}}/4$. The state with all $| xy \rangle$ occupied
is degenerate with the state with all $| yz \rangle$ occupied, see
Fig. 4. Those states do differ, however, because the dimerization
pattern along the chain is shifted by one lattice spacing. At zero
temperature the system is condensed in either one of these two
dimerized orbital ordered states and the translation symmetry is
broken. This explains the structural change at $T_c$~\cite{a7},
the symmetry change and the energy shifts of the phonon
excitations in the Raman spectra of NaTiSi$_2$O$_6$, and the
observation of a large spin gap in the susceptibility measurements
~\cite{a2}.

At high temperatures, due to strong orbital-lattice coupling and
JT effect, the orbital fluctuations produce dramatic effects.
Above T$_{\rm OO}$ the JT distortion disappears: the high-T
orbital disordered phase of NaTiSi$_2$O$_6$ may be regarded as an
orbital fluctuating phase, and on larger time scales the TiO$_6$
octahedra appear to be undistorted. This causes the crystal to be
effectively more symmetric, which is in agreement with the
observation of less-then-expected phonon modes in the room
temperature Raman spectra. As a consequence, orbital fluctuations
induce large phonon broadenings (of course, the modes with Ti-O
bonds in their normal coordinates will be mostly affected), as
they are indeed found in the room temperature Raman spectra of
NaTiSi$_2$O$_6$, see Fig. 2. In that respect the high-T phase of
NaTiSi$_2$O$_6$ resembles a {\it dynamical JT phase} (where the
phonon broadening is a signature of the melted static lattice
distortions~\cite{a771,a772}). The spin fluctuations above the
orbital ordering temperature \cite{a2} can be easily understood in
a straightforward mean-field approximation of
Hamiltonian~(\ref{eq:ham}), where we can decouple orbitals and
spins. If the long range order is absent in the orbital sector, we
have for the expectation values $\langle T^z_i \rangle =0$, and
$|\langle T^z_i T^z_j \rangle | < 1/4 $, so that the effective
exchange constants --and therefore the fluctuations in the
spin-sector-- are still antiferromagnetic. Thus, the nature of the
phase transition in NaTiSi$_2$O$_6$ can be established: it
corresponds to an orbital order-disorder phase transition with
appropriate concomitant magnetic and lattice changes.

In the case of two electrons per site (NaVSi$_2$O$_6$), the
orbital fluctuations are strongly suppressed due to inert property
of $| xz \rangle$ orbitals, and the corresponding phonon
broadenings are much smaller, see Fig. 2. NaCrSi$_2$O$_6$ has a
fully polarized t$_{2g}$ core, no orbital degrees of freedom, and
no anomalous phonon broadenings in the Raman spectra.

{\it Conclusions.} We report a study of the two types of the phase
transitions, observed in pyroxene family, by analyzing the
Raman-active phonon and magnon excitations, and their temperature
dependence. We find that spin S=1/2 compound, NaTiSi$_2$O$_6$,
exhibits a transition at about 210 K that we assign to be an orbital
order-disorder phase transition. It originates from the instability
of the high temperature orbital fluctuating - dynamical Jahn-Teller
phase, towards a dimerized orbital ordered state, which is accompanied
by a lattice distortion and by spin valence bond formation.
The spin S=1 system, NaVSi$_2$O$_6$, on the contrary, does not show
this type of instability and orders as a N\`eel-type antiferromagnet
below T$_N$ = 19 K in agreement with our microscopic spin-orbital
model for the pyroxenes.

M.J.K. and Z.V.P. acknowledge support from the Research Council of
the K.U. Leuven and DWTC.  The work at the K.U. Leuven is
supported by Belgian IUAP and Flemish FWO and GOA Programs.

*milan.konstantinovic@fys.kuleuven.ac.be

**Present address: Materials Science Institut, University of
Valencia, Poligono La Coma, 46980 Paterna (Valencia), Spain}


\begin{thebibliography}{99}
\bibitem{a701} The splitting of $t_{2g}$'s due to non-cubic crystal fields
is generally weak and smaller than typical hopping integrals.
\bibitem{a5b} We neglect the weak ferromangetism that results from the Hund's rule exchange.
\end{thebibliography}
\end{document}